\documentclass[nofootinbib, floatfix, prd, aps, twocolumn, superscriptaddress, preprintnumbers]{revtex4-2}

\usepackage{relsize}

\usepackage[acronyms, nohypertypes={acronym}, nopostdot, style=super, nonumberlist, toc]{glossaries}
\glsenableentrycount
\setacronymstyle{long-sc-short}

\newcommand\ac[1]{\gls{#1}}

\def\acresetall{\glsresetall[acronym]}

\newacronym{WF}{wf}{Wilson-Fisher}
\newacronym{AF}{af}{asymptotically free}
\newacronym{RG}{rg}{renormalization group}
\newacronym{WZW}{wzw}{Wess-Zumino-Witten}
\newacronym[longplural={conformal field theories}]{CFT}{cft}{conformal field theory}
\newacronym[longplural={lattice field theories}]{LFT}{lft}{lattice field theory}
\newacronym[longplural={effective field theories}]{EFT}{eft}{effective field theory}
\newacronym[longplural={quantum field theories}]{QFT}{qft}{quantum field theory}
\newacronym{LEC}{lec}{low-energy constant}
\newacronym{QCD}{qcd}{quantum chromodynamics}
\newacronym{MC}{mc}{Monte Carlo}
\newacronym{IR}{ir}{infrared}
\newacronym{UV}{uv}{ultraviolet}
\newacronym{SNR}{snr}{signal-to-noise ratio}
\newacronym{NLSM}{nl$\sigma$m}{nonlinear sigma model}
\newacronym{PCM}{pcm}{principal chiral model}
\newacronym{CSA}{csa}{Cartan subalgebra}
\newacronym{SSB}{ssb}{spontaneous symmetry breaking}
\newacronym{DOF}{dof}{degrees of freedom}
\newacronym{DMRG}{dmrg}{densiy matrix renormalization group}
\newacronym{YM}{ym}{Yang-Mills}
\newacronym{QLM}{qlm}{quantum link model}
\newacronym{KG}{kg}{Kogut-Susskind}
\newacronym{KG-QLM}{kg-qlm}{Kogut-Susskind quantum link model}
\newacronym{D-QLM}{d-qlm}{D-theory quantum link model}
\newacronym{SPT}{spt}{symmetry-protected topological}
\newacronym{GW}{gw}{Ginsparg-Wilson}
\newacronym{FGW}{fgw}{Fujikawa-Ginsparg-Wilson}
\newacronym{FK}{fk}{Fidkowski-Kitaev}
\newacronym{CS}{cs}{Chern-Simons}
\newacronym{APS}{aps}{Atiyah-Patodi-Singer}
\newacronym{PV}{pv}{Pauli-Villars}
\newacronym{PBC}{pbc}{periodic boundary conditions}
\newacronym{OBC}{obc}{open boundary conditions}
\newacronym{ABC}{abc}{antiperiodic boundary conditions}
\newacronym{KD}{kd}{Kahler-Dirac}
\newacronym{RKD}{rkd}{reduced Kahler-Dirac}
\newacronym{SMG}{smg}{symmetric mass generation}

\newacronym{CHN}{chn}{Creutz-Horvath-Neuberger}
\newacronym{GEP}{gep}{generalized eigenvalue problem}

\usepackage{mathtools}
\usepackage{amsmath,amssymb,amsbsy,amsfonts}

\usepackage{graphics}
\graphicspath{{./fig/}}

\usepackage[normalem]{ulem}

\usepackage{physics}
\usepackage{physics2}
\usephysicsmodule{ab,ab.braket}

\usepackage{tikz}

\renewcommand\>{\rangle}
\newcommand\<{\langle}

\newcommand\Order{O}

\usepackage{microtype}

\newcommand\del\partial

\usepackage{dutchcal}

\newcommand\tsum{\textstyle\sum}

\usepackage{xcolor}
\usepackage{colortbl}
\usepackage{tabularx}

\usepackage{makecell}

\newcommand\minisec[1]{\emph{#1}---}

\usepackage{enumitem}
\usepackage{pifont}

\usepackage{graphicx}
\usepackage{amsfonts}
\usepackage{tocvsec2}
\usepackage{slashed}
\usepackage{cancel}
\usepackage{comment}

\usepackage{minibox}
\usepackage{orcidlink}
\usepackage{lipsum}

\newcommand\beq{\begin{eqnarray}}
\newcommand\eeq{\end{eqnarray}}

\newcommand\sign{\mathrm{sign}}

\usepackage{xparse}

\NewDocumentCommand{\Dw}{o}{%
  \IfValueTF{#1}{%
    D^{(#1)}_W
  }{%
    D_W
  }
}

\newcommand\Qa{{Q}_5}
\newcommand\gmodDef{\hat{\gamma}_5}

\NewDocumentCommand{\Hov}{o}{\IfValueTF{#1}{\hat{h}^{#1}}{\hat{h}}}
\NewDocumentCommand{\hov}{o}{\IfValueTF{#1}{h_{(#1)}}{h}}
\NewDocumentCommand{\Dov}{o}{\IfValueTF{#1}{D_{(#1)}}{D}}
\NewDocumentCommand{\Vov}{o}{\IfValueTF{#1}{V_{#1}}{V}}

\NewDocumentCommand{\gmod}{o}{\IfValueTF{#1}{\hat{\gamma}_{5(#1)}}{\gmodDef}}

\newcommand\Pbdy{P_{\text{low}}}
\newcommand\Pbulk{P_{\text{high}}}
\newcommand\xs{x_5}
\newcommand\Ls{L_5}
\newcommand\Hdw{H_{\text{DW}}}
\newcommand\hdw{h_{\text{DW}}}

\usepackage{ninecolors}
\usepackage{hyperref}
\hypersetup{
  unicode,
  pdfprintscaling=None,
  breaklinks,
  colorlinks=true,
  linkcolor=azure5,
  citecolor=azure5,
  filecolor=magenta,
  urlcolor=blue5,
}
\usepackage[capitalise]{cleveref}

\renewcommand\Ls{\mathcal{l}}
\newcommand\Lbulk{L_5}
\newcommand\jchi{j_5}
\newcommand\Canom{\mathcal{C}}
\newcommand\deltasl{\slashed{\delta}}
\newcommand\dsh{\hat{\slashed{\delta}}}

\begin{document}

\preprint{FERMILAB-PUB-25-0330-T}

\title{Ginsparg-Wilson Hamiltonians with Improved Chiral Symmetry}

\author{Hersh Singh\,\orcidlink{0000-0002-2002-6959}}
\email{hershs@fnal.gov}
\affiliation{Fermi National Accelerator Laboratory, Batavia, Illinois 60510, USA}

\begin{abstract}
  We construct a family of Ginsparg-Wilson Hamiltonians with improved chiral properties, starting from a construction of Creutz-Horvath-Neuberger that provides a doubler-free Hamiltonian lattice regularization for Dirac fermions in even spacetime dimensions. We use a higher-order generalization of the \ac{GW} relation due to Fujikawa, which yields an order-$k$ Hamiltonian overlap operator for each integer $k \geq 0$, with an exactly conserved but nonquantized chiral charge that becomes quantized as $k \to \infty$. Our construction provides physical insight into how Fujikawa's higher-order \ac{GW} relation improves chiral symmetry while reproducing the anomaly, highlighting the trade-offs inherent in any Hamiltonian lattice realization of an anomalous chiral symmetry. This class of Hamiltonian lattice regularizations, with their tunable chiral symmetry properties, offers potential advantages for quantum and tensor-network simulations.
\end{abstract}

\maketitle
\acresetall

\section{Introduction}

A practical nonperturbative construction of chiral gauge theories remains an outstanding problem \cite{kaplan_chiral_2012, nielsen_nogo_1981, nielsen_absence_1981a, nielsen_absence_1981}.
The \ac{GW} relation \cite{ginsparg_remnant_1982} and its solutions \cite{kaplan_method_1992,neuberger_exactly_1998,neuberger_more_1998,hasenfratz_prospects_1998}
have greatly improved our understanding of chiral symmetry on the lattice.
\ac{GW} fermions admit an exact modified chiral symmetry \cite{luscher_exact_1998}, which Lüscher used to construct lattice abelian chiral gauge theories \cite{luscher_abelian_1999}.
For nonabelian theories, the problem remains open despite \cite{kikukawa_domain_2002, luscher_lattice_2000} the recent resurgence of activity \cite{aoki_lattice_2024a, aoki_lattice_2024b, berkowitz_exact_2023, catterall_chiral_2021, catterall_lattice_2024, kaplan_chiral_2024, kaplan_weyl_2024, pedersen_reformulation_2023, wang_nonperturbative_2022, wang_solution_2019, wang_symmetric_2022a, zeng_symmetric_2022, golterman_conserved_2024, golterman_propagator_2024, clancy_generalized_2024, clancy_ginspargwilson_2024,kan_lattice_2025}.

The overlap operator \cite{neuberger_more_1998,neuberger_exactly_1998} provides an elegant closed-form solution to the \ac{GW} relation.
However, much of the work involving \ac{GW} or overlap fermions has been conducted within the Euclidean path integral formulation.
A Hamiltonian construction is desirable for several reasons.
Not only does it often provide deeper physical intuition, but Hamiltonian formulations are also essential for enabling quantum computing and tensor network methods, which are likely crucial due to the expected sign problems in chiral gauge theory.
Finally, a Hamiltonian formulation establishes direct contact with the growing condensed matter literature that links anomalies and topological phases.

A Hamiltonian \ac{GW} operator has occasionally been pursued \cite{ horvath_ginspargwilson_1999,cheluvaraja_axial_2001,creutz_new_2002,clancy_ginspargwilson_2024}.
Refs.~\cite{cheluvaraja_axial_2001,creutz_new_2002}, in particular, proposed a definition of an overlap Hamiltonian, which we call the \emph{standard} overlap Hamiltonian.
This admits an exact (modified) chiral charge operator, analogous to the Euclidean overlap operator \cite{luscher_exact_1998}.
Although the modified chiral charge yields a compact $U(1)$ chiral symmetry in Euclidean spacetime, the overlap Hamiltonian is fundamentally different: the exact chiral charge has a non-quantized spectrum and therefore generates a non-compact $U(1)$ symmetry on the lattice.
Exact chiral symmetries and the challenges of formulating a quantized charge in Hamiltonian treatments of Dirac fermions have recently received attention, as there is a tension between locality, unitarity, and the compactness of the $U(1)$ chiral symmetry
\cite{clancy_ginspargwilson_2024,mandula_symmetries_2009,clancy_generalized_2024, chatterjee_quantized_2024, gioia_exact_2025, yamaoka_quantized_2025}.
This difficulty is consistent with a no-go theorem of Fidkowski and Xu \cite{fidkowski_nogo_2023}, which prohibits a local quantized chiral charge for Weyl fermions on the lattice.
However, from a continuum perspective, the absence of a quantized chiral charge seems unsatisfactory.
If a Hamiltonian overlap with a compact $U(1)$ chiral symmetry is indeed impossible, we must ask: Can the quantization of the chiral charge be systematically improved, and what insights might emerge from a formulation of a Hamiltonian overlap operator?

In this work, we construct a family of overlap Hamiltonians (in arbitrary odd spatial dimensions) indexed by a positive odd integer $k$.
We do this using an algebraic generalization of the \ac{GW} relation due to Fujikawa \cite{fujikawa_algebraic_2000, fujikawa_aspects_2001}.
We find that this approach yields an infinite tower of $k$-overlap Hamiltonians
with exactly conserved chiral charges.
We study the spectrum of these $k$-overlap Hamiltonians in $1+1$ dimensions and demonstrate how the quantization of chiral charges improves with increasing $k$.
We show how this construction clarifies the role of the anomaly in enforcing a trade-off between locality and the quantization of the chiral charge.
Finally, we argue that the lack of quantization and its improvement have a straightforward physical interpretation when viewed as the boundary theory of a bulk \ac{SPT} phase, as in domain-wall fermions.\footnote{
    \emph{Notation:}
    We use the mostly-minus signature in $(d+1)$ space-time dimensions $\eta^{\mu \nu} = \operatorname{diag}(1, -1, \dotsc, -1)$.
    The Dirac $\gamma$ matrices satisfying $\ab\{ \gamma^{\mu}, \gamma^{\nu} \} = 2 \eta^{\mu \nu}$.
    We take $\gamma_0, \gamma_{5}$ to be Hermitian and $\gamma_{a}$ $(a=1, \dotsc, d)$ to be anti-Hermitian.
    When writing fermionic Hamiltonians such as $H = {\psi}^{\dagger} h \psi$, we suppress the sum over any spatial and spinor indices.
}

\section{Higher-order Ginsparg-Wilson Hamiltonians}
\label{sec:fujikawa}

\minisec{Standard Ginsparg-Wilson Hamiltonian}
Let us recall the standard construction of a Hamiltonian \ac{GW}  operator, following \ac{CHN} \cite{creutz_new_2002}.
In $d$ (odd) spatial dimensions, let $\psi_{x}$ denote a $2^{\frac{d+1}{2}}$-component lattice Dirac fermion field at the lattice site $x$.
The standard \ac{GW} fermion Hamiltonian is:
\begin{align}
	H =  \psi^{\dagger} h \psi, \quad h = \gamma_{0} D .
      \label{eq:chn}
\end{align}
where $h = \gamma^{0} D$ is the single-particle Hamiltonian, with $D$ being a \emph{spatial} \ac{GW}-Dirac operator.
The operator $D$ satisfies the \ac{GW} relation
\begin{align}
  D + {D}^{\dagger} &= 2 {D}^{\dagger} D 
  \label{eq:gw}
\end{align}
where the lattice spacing has been set to $a=1$,
along with $\gamma_{5}$ and $\gamma_{0}$ hermiticity,
\begin{align}
  \gamma_{5} D \gamma_{5} = \gamma_0 D \gamma_0 = {D}^{\dagger}.
\end{align}
In particular, $D$ may be taken as the $d$-dimensional overlap operator \cite{neuberger_exactly_1998,neuberger_more_1998}.
Crucially, this Hamiltonian has an exactly conserved (modified) chiral charge \cite{creutz_new_2002}, akin to a Euclidean Ginsparg-Wilson operator \cite{luscher_exact_1998}.
We define a “modified” $\gamma_{5}$:
\begin{align}
  \gmod &= \gamma_5 ( 1 - D ),
              \label{eq:chn-charge}
\end{align}
which reduces to $\gamma_{5}$ in the continuum limit $D \to 0$.
This allows us to define a (second-quantized) chiral charge operator $\Qa = {\psi}^{\dagger} \gmod \psi$.
The \ac{GW} relation implies that $[\gmod, \hov] = 0$, and thus $[\Qa, H] = 0$.
The single-particle Hamiltonian $\hov$ and the single-particle chiral charge $\gmod$ are simply related:
\begin{align}
	\gmod^{2} + \hov^{2} &= 1.
                        \label{eq:qh0}
\end{align}
\Cref{eq:qh0} then implies that the $(\gmod, \hov)$ eigenvalues lie on a circle.
In particular, low-energy states with $h \sim 0$ have a well-defined chiral charge $\gmod \sim \pm 1$, while high-energy states $h \sim 1$ have a vanishing chiral charge $\gmod \sim 0$.
This clearly shows that the chiral charge $\Qa$ is not quantized and thus does not generate a compact $U(1)$ symmetry on the lattice.
Can this be improved within a \ac{GW} framework?

\minisec{Improved chiral symmetry}
To improve the chiral symmetry properties of the \emph{Euclidean} overlap operator,
Fujikawa \cite{fujikawa_algebraic_2000, fujikawa_aspects_2001} proposed a family of higher-order \ac{GW} relations.
In a form analogous to \cref{eq:gw}, these order-$k$ \ac{FGW} relations can be written as
\begin{align}
  D + {D}^{\dagger} &= 2 ({D}^{\dagger} D)^{k+1},
                     \label{eq:fgw}
\end{align}
where $k \geq 0$ is an integer.  For $k=0$, this is the ordinary \ac{GW} relation \eqref{eq:gw}.
The higher powers of ${D}^{\dagger} D$ on the right-hand side suggest a Symanzik-type improvement of chiral symmetry.
The Euclidean generalized \ac{GW} relations admit overlap-like solutions with no doublers, which reproduce the anomaly correctly and are thus valid regularizations of a Dirac fermion
\cite{fujikawa_chiral_2000, fujikawa_generalized_2002}.

\begin{figure*}[!ht]
  \centering
  \includegraphics[width = 0.99\linewidth]{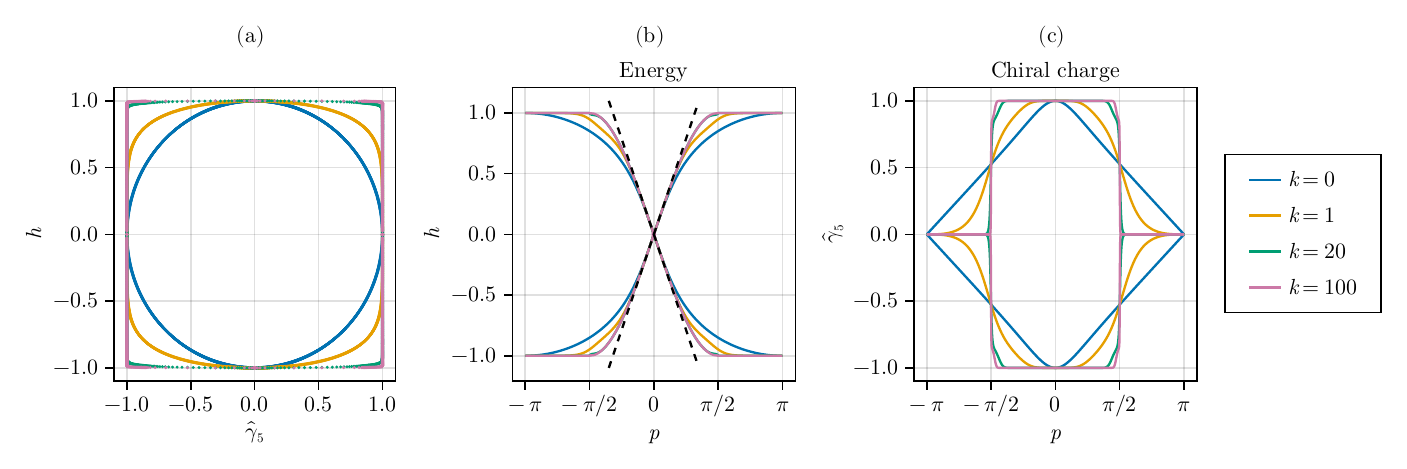}
  \caption{
    Behavior of order-$k$ Hamiltonian overlap operators in $d=1$ spatial dimensions on a $L$-site lattice with antiperiodic boundary conditions for varying $k$.
    (a) The relationship in \cref{eq:energy-charge} between energies and chiral charge, independent of dimension $d$.
    The standard ($k=1$) overlap traces a circle on this plot, while the higher-$k$ approach a rectangular shape, improving the quantization.
    (b) Spectrum of $h(p)$ as a function of momentum $p$.
    As $k$ becomes large, the dispersion relation for $|p| < \frac{\pi}{2}$ approaches the naive discretization, while the higher-modes are pushed to $|h| = 1 $.
    (c) Spectrum of the single-particle chiral charge operator $\gmod(p)$.
    The chiral charge for the standard overlap ($k=1$) falls linearly with $|p|$.
    However, higher $k$ have a sharper fall-off near $|p| = \pi / 2$.
    As we increase $k$, the modes below $|p| < \frac{\pi}{2}$ are pushed toward $|\gmod| = 1$, while all the high-momentum modes outside this region are pushed towards zero chiral charge.
    In the limit $k \to \infty$, this becomes exact with the modified chiral charge taking values exactly $\pm 1$ for the low-momentum modes ($|p| <  \pi / 2 $) and vanishing for the high-momentum modes $(|p| > \pi / 2)$.
    \label{fig:gwk}}
\end{figure*}

Now, we replace the conventional $k=0$ \ac{GW} relation in the standard Hamiltonian \ac{GW} construction [\cref{eq:chn}] with the higher-order \ac{FGW} relation of \cref{eq:fgw}.
We can again write an exactly conserved chiral charge $\Qa = {\psi}^{\dagger} \gmod \psi$, with a modified $\gamma_{5}$:
\begin{align}
  \gmod &= \gamma_{5} - (\gamma_5 \Dov)^{2k+1},
\end{align}
which generalizes \cref{eq:chn-charge}.
How is the relation in \cref{eq:qh0} between energy and chiral charge of the single-particle modes modified by this generalization?
Interestingly, we find that
\begin{align}
	\gmod^2 + \hov^{4k+2} = 1.
  \label{eq:energy-charge}
\end{align}
We show the behavior of $(\gmod, \hov)$ eigenvalues in \cref{fig:gwk}(a) for various $k$.
In particular, as $k \to \infty$, the operator $\gmod$ becomes increasingly quantized with $|\gmod| = 1$ for any mode with $|\hov| < 1$ (``finite-energy'' modes) or $|\gmod| = 0$ for modes with $|\hov| = 1$ (``infinite-energy'' modes).
For large but finite $k$, the violations of quantization come only from high-energy modes where the chiral charge changes rapidly.
We will discuss a physical interpretation of this below.

\minisec{Order-$k$ overlap Hamiltonians}
To complete the construction, we need an explicit form for a Hamiltonian overlap operator that satisfies the order-$k$ \ac{FGW} relation, which we refer to as the $k$-overlap Hamiltonian.
Fortunately, the Euclidean order-$k$ overlap operators \cite{fujikawa_generalized_2002,fujikawa_locality_2001,fujikawa_algebraic_2000,fujikawa_chiral_2000,fujikawa_properties_2002} can be adapted for this purpose.
We first define
\begin{align}
	\Dw[n] &= i (\hat\deltasl)^{n} - \ab( r \Delta / 2 )^{n} - m^{n}
           \label{eq:Dw-n}
\end{align}
where
$\hat\delta_{a} = \frac{1}{2} \ab(\delta_a - \delta_a^{\dagger})$ and
$\Delta = - \sum_{a=1}^{d} {\delta}^{\dagger}_{a} \delta_{a}$, with
$\delta_{a}$ being the forward-difference operator acting as
$ \delta_{a} \psi_{x} =   \psi_{x + \hat{a}} - \psi_{x}$
for $\hat{a}$ a unit vector in the spatial direction $a$.
For $n=1$, this is the standard Wilson-Dirac operator with the parameter $m,r$.
Finally,  defining
\begin{align}
  \hov[2k+1] &= \frac{1}{2} \gamma^0  \ab(1 + \varepsilon\ab[ \Dw[2k+1] ]),
\end{align}
with $\varepsilon(X) = X / \sqrt{{X}^{\dagger} X}$,
the $k$-overlap Hamiltonian is given by
$h = \ab[\hov[2k+1]]^{\frac{1}{2k+1}}$.
As in the standard overlap, we set $0 < m < 2r$ to avoid doublers and further require $2m^{2k+1} = 1$ for normalization.
More details of this construction, along with a prescription for the $(2k+1)$th root, are given in \cref{sec:explicit}.

\minisec{1+1 dimensions}
We can gain insight into the $k$-overlap Hamiltonians by studying them in $(1+1)$ dimensions, where the behavior becomes particularly simple.
In \cref{fig:gwk}, we show the behavior of the single-particle modes of the order-$k$ overlap operators on a periodic $L$-site lattice.
The first panel shows the relationship between the single-particle energies $h$ and the (modified) chiral charge $\gmod$ predicted by \cref{eq:energy-charge}.
The standard overlap Hamiltonian traces a circle on this plot, whereas the higher $k$-overlap Hamiltonians approach a rectangular shape.
This is how Fujikawa's higher-order \ac{GW} relations improve chiral charge quantization --- the low-energy modes are pushed closer to the edges $\gmod = \pm 1$.

\cref{fig:gwk}(b,c) show the single-particle energy and chiral charge as functions of momentum $p$ for various $k$.
The linear low-energy dispersion relation $h = \pm p$ is preserved for low momentum $|p|\sim 0 $ as we increase $k$, but higher momentum modes ($|p| \gtrsim \frac{\pi}{2}$) are systematically pushed to the cutoff $|h| = 1$.
Similarly, the chiral charge is always $\pm 1$ near $|p| \sim 0$, but it vanishes at high momentum.
Increasing $k$ pushes all $|p| > \frac{\pi}{2}$ modes toward $|h| = 1$ and $\gmod = 0$.

\minisec{The $k \to \infty$ limit}
The above discussion suggests that the $k \to \infty$ overlap Hamiltonian is particularly simple.
Indeed, if we define projectors $\Pbdy$ for $|p| < \frac{\pi}{2}$ modes, and $\Pbulk$ for the remaining modes, the Hamiltonian and the chiral charge in the $k \to \infty$ limit are:
\begin{align}
  (h)_{\infty} &= \gamma_0\gamma_1 \sin p_{1} \Pbdy  + \gamma_0 \Pbulk.\\
  (\gmod)_{\infty} &= \gamma_5 \Pbdy.
\end{align}
We find that the $k \to \infty$ overlap Hamiltonian takes the naive discretization $h(p) = \gamma_0 \gamma_1 \sin p_{1}$ of the continuum Dirac Hamiltonian $h = \gamma_0 \slashed{\del} $ and simply lifts all high-momentum modes to $|h| = 1$, while the low-momentum modes $|p| < \frac{\pi}{2}$ remain unaffected.\footnote{
  In higher-dimensions, we find a similar behavior. There is a low-momentum region of the Brillouin zone where the naive discretization is left untouched while all the modes outside it get lifted to high energies.
  However, the shape of this region is more complicated than in $d=1$.}

This also shows how nonlocality in the Hamiltonian and the chiral charge operator arises in the $k \to \infty$ limit.
When $k \to \infty$, the second derivative of $h(p)$ becomes discontinuous at $|p| = \frac{\pi}{2}$, even though $h(p)$ and its first derivative are continuous.
In the case without gauge fields, we can trivially discard the $\Pbulk$ modes, resulting in an ultralocal, quantized chiral charge that commutes with the $U(1)$ vector charge.
However, the nonanalyticity
at $p = \frac{\pi}{2}$ makes the Hamiltonian nonlocal.\footnote{Such a formulation in the $k \to \infty$ limit is effectively the same as SLAC fermions \cite{drell_strong_1976} or tangent/Stacey fermions \cite{stacey_eliminating_1982} studied recently in Ref.~\cite{haegeman_interacting_2024}, both of which have an ultralocal chiral charge but a nonlocal Hamiltonian.}
This behavior persists in higher dimensions as well.
In fact, it can be shown that in arbitrary dimensions \cite{fujikawa_locality_2001,fujikawa_properties_2002} ,
\begin{align}
	(\gamma_5 \Dov)^{2k+1} \sim e^{- |x-y| / k}.
  \label{eq:Hchi-decay}
\end{align}
Therefore, the charge operator $\gmod  = \gamma_5 - (\gamma_5 \Dov )^{2k+1}$ is exponentially local (but unquantized) for any finite value of $k$.
In the $k \to \infty$ limit, it becomes quantized but ceases to be exponentially local.
In this manner, we can trade locality for chirality by increasing $k$.

\section{Chiral anomaly and the domain-wall spectrum}
\label{sec:anomaly}

\begin{figure*}[htbp]
  \centering
  \raisebox{-0.5\height}{\includegraphics[width=0.99\linewidth]{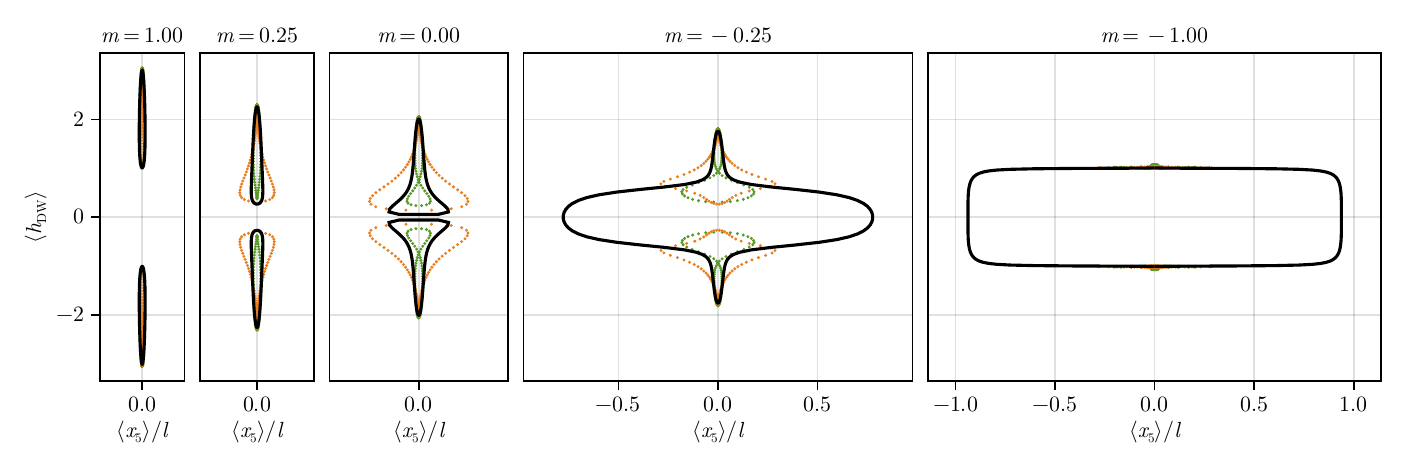}}
  \caption{
    \label{fig:edgemodes}
      Spectrum of the domain-wall Hamiltonian $\hdw(\vec p)$ in 2 spatial dimensions
      shown against the (normalized) mean position $\< \xs \> / \Ls$ along the extra dimension, as
      $m$ is tuned from $m > 0$  (trivial phase) to $m < 0$ (nontrivial \ac{SPT} phase).
      We show the lowest few bands, highlighting two bands by a solid black line.
      For $m > 0$, all bands are concentrated in the center.
      As $m \to 0^{+}$, the bands widen, signaling a second-order phase transition.
      At $m=0$, the black bands with $\hdw > 0 $ and $\hdw < 0$ merge causing the appearance of zero-modes.
      For $m < 0$, these chiral zero-modes become increasingly localized at the edges $\braket< \xs > /l = \pm 1$ as we decrease $m$ further, while the other bands become heavy again and return into the bulk.
      Taking the displacement from the center as a measure of the boundary chiral charge shows how the $k$-overlap Hamiltonians elegantly capture the essential low-energy physics, detemined by the black band for $m<0$.
  }

\end{figure*}

The Euclidean overlap operator can be derived as the effective boundary operator for domain-wall fermions  \cite{narayanan_chiral_1994,kikukawa_low_2000a}. Domain-wall fermions \cite{kaplan_method_1992} provide a clear understanding of the chiral anomaly based on the anomaly inflow \cite{callan_anomalies_1985}.
In this framework, the nonconservation of chiral charge for the $d$-dimensional boundary fermions in the presence of $U(1)$ gauge fields is facilitated by a current flowing into the $(d+1)$-dimensional bulk.

Does the correspondence between overlap and domain-wall formulations continue to hold in the Hamiltonian formulation? If so, can we gain deeper insight into the issue of quantization?
We have already seen that the $k$-overlap Hamiltonians offer a way to improve the quantization of the chiral charge at the expense of locality.
In this section, we discuss how anomaly inflow \cite{callan_anomalies_1985} provides a physical understanding of why we encounter obstacles in constructing a local, quantized chiral charge that commutes with the vector $U(1)$ charge.

Let us first recall how the chiral anomaly arises in the Hamiltonian domain-wall formalism \cite{creutz_surface_1994}.
The domain-wall Hamiltonian $\Hdw = \psi^\dagger \hdw \psi$ is defined on a $(d+1)$ (spatial) dimensional lattice
with \ac{OBC} along the bulk direction, with an extent of $\Lbulk = 2\Ls + 1$, and \ac{PBC} along the $d$-dimensional boundary.
The single-particle domain-wall Hamiltonian is $\hdw = \gamma^0 \Dw[1]$, where $\Dw[1]$ is the standard Wilson-Dirac operator from \cref{eq:Dw-n} with $r = 1$ and a mass parameter $m$.
Let $\vec p$ be the momentum along the boundary, and $\xs \in [-\Ls, \Ls]$ be the coordinate along the extra dimension.
In momentum space along the boundary, the domain-wall Hamiltonian is given by
$\gamma^0\hdw(\vec p) =  i \gamma_5 \hat\delta_{5} - \frac{r}{2} \Delta_5 +  m + \sum_{{a=1}}^d \ab[  \gamma^{a} \sin(p_a) + r \ab(1-\cos p_a) ]$.

In \cref{fig:edgemodes}, we show how the spectrum of $\hdw(\vec p)$ in $d+1=2$ (spatial) bulk dimensions varies with $m$.
For each eigenstate $\ket| \lambda > $ of $\hdw(\vec p)$, we plot the energy $ \< \lambda | \hdw(\vec p) | \lambda \>$ against the position along the extra dimension $\< \lambda | \xs | \lambda \>$.
Varying $\vec p$ and tracking the eigenvalues results in the bands in \cref{fig:edgemodes}.
For $m > 0$, we note that all the modes are heavy and clustered deep within the bulk $ \xs \sim 0$.
As $m$ passes through a phase transition at $m=0$, we observe the emergence of massless modes appearing at the boundaries.
These are the chiral edge modes.

\begin{figure*}[htbp]
  \centering
  \includegraphics[width = \linewidth]{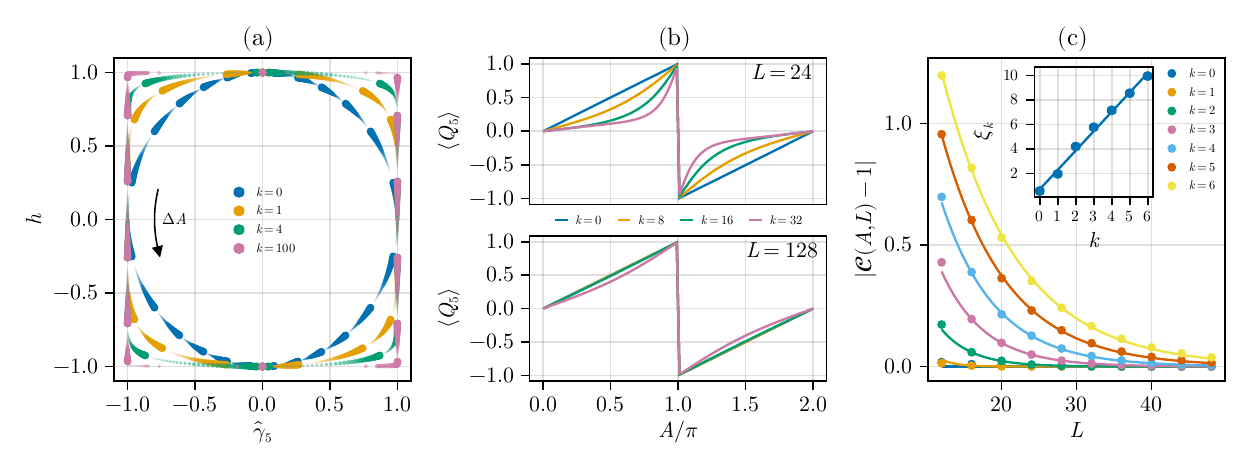}
  \caption{
    Anomaly inflow and finite-size effects for the $k$-overlap operators in $d=1$ spatial dimension with a varying background gauge field.
    (a)
    Flow of single-particle modes in the $(h, \gmod)$ plane as $A \to A + 2\pi$.
    As $A$ is varied, $\gmod=-1$ modes smoothly turn into $\gmod = +1$ by passing through $\gmod = 0$.
    This is the same picture as domain-wall fermions \cite{creutz_surface_1994} if we identify $\gmod$ as the coordinate along the extra dimension.
    Modes for higher $k$ have faster speeds when leaving the ``walls'' at $|\gmod| = 1 $.
    (b)
    The vacuum chiral charge $\< \Qa \>$ as a function of $A$ with varying $k$ for $L=24$ (above) and $L=128$ (below).
    As we increase $k$, the finite volume effects become stronger.
    The continuum prediction of a linear behavior is achieved by the $k=0$ even for $L=24$, but the higher $k$ curves become linear only with a larger value of $L = 128$.
    (c)
    The deviation from the continuum result $\mathcal{C}$ in \cref{eq:anomaly-C} as a function of $L$ for various values of $k$.
    Larger $k$ require a larger $L$: the convergence for a fixed $k$ is exponential in $L$, but the decay length is proportional to $k$, as shown in the inset with $\xi_{k} \sim 1.6k$.
    \label{fig:anomaly-inflow}
  }
\end{figure*}

Note that the relationship between energy $\braket< \hdw >$ and bulk position $\< \xs \>$, as shown by the black band in \cref{fig:edgemodes} for the $m < 0$ regime, closely resembles the relationship between $\hov$ and $\gmod$ depicted in \cref{fig:gwk}(a) for the $k$-overlap operators.
Indeed, as noted in Ref.~\cite{creutz_surface_1994}, a natural measure of the chiral charge of the boundary modes is simply the normalized displacement from the center: $\Qa = \< \xs \> / \Ls$.
This is clearly an unquantized definition.
The low-energy modes have $\Qa \sim \pm 1$, while the high-energy ``bulk'' modes have $\Qa \sim 0$.
We see that an unquantized charge is very natural when considering the anomalous Dirac fermion as the boundary of a higher-dimensional bulk theory — the high-energy bulk modes have no notion of chirality, so their chiral charge is zero.

This picture also provides physical intuition for what it means to improve chiral symmetry. The violation of the compactness of the $U(1)$ chiral symmetry arises from the high-energy modes moving away from the wall into the bulk.
To improve this, we could ``push'' the modes closer to the wall.
But this is exactly what \cref{eq:energy-charge} does: $\gmod^2 + h^{4k+2} = 1$.
As $k$ increases, all low-energy modes are pushed to the ``walls'' at $\gmod \sim \pm 1$.
However, as we improve the chiral properties at low energies, the bulk modes get pushed deeper into the bulk $\gmod = 0$.
In the $k \to \infty$ limit, there is an extensive number of such bulk modes with $\gmod = 0$.

\minisec{Chiral anomaly}
We have identified the $\gmod = 0$ modes in the overlap Hamiltonian as effective bulk modes.
But is it essential to retain them in a Hamiltonian formulation of the boundary theory?
Here, we provide intuition for why these modes must necessarily exist—they ensure anomaly inflow.

The anomaly inflow mechanism is particularly transparent in $1+1$ (boundary) dimensions.
The overlap Hamiltonians allow for straightforward background gauging of the (onsite) $U(1)$ vector symmetry.
(The analysis for the domain-wall Hamiltonian is identical once we identify $\Qa \sim \< \xs \> / \Ls $, as argued above.)
With background gauge fields $A_{\mu}$ ($\mu = 0,1$),
the continuum prediction for the anomalous nonconservation of the chiral current $\jchi^{\mu}$ states
$ \braket< \del_{\mu} \jchi^{\mu} > = \frac{1}{2\pi} \epsilon_{\mu \nu} \braket< F^{\mu \nu} >$, where $F_{\mu\nu} = \del_{\mu} A_{\nu} - \del_{\nu} A_{\mu}$.
In the gauge $A_{0}=0$ and $ \del_{1} A_{1} = 0$,
the gauge field is described by a spatially constant (but possibly time-varying) $A_1 \equiv A(t) \in [0, 2\pi / L)$.
Integrating over space, we obtain
\begin{align}
  \braket< \Qa(A) > - \braket<\Qa(0) > &= A L / \pi.
                     \label{eq:Q(A)}
\end{align}
The momentum-space overlap Hamiltonian with a background gauge field $A$ is given by $\hov(\vec p + \vec A)$ for a Dirac fermion with unit $U(1)$ vector charge.

In \cref{fig:anomaly-inflow}(a), we show how the modes of the $k$-overlap Hamiltonian move within the $(h, \gmod)$ space as we turn on the background gauge field $A$ for various values of $k$.
As we increase the background gauge field $A$, the eigenvalues move along the curve defined by \cref{eq:energy-charge}.
The modes localized on the left wall $\gmod = -1$ increase in energy and migrate into the bulk $\gmod \sim 0$. 
Simultaneously, the bulk modes shift towards the right wall $\gmod = +1$.
This is how the chiral anomaly arises: for left-handed modes to smoothly transform into right-handed modes, they must traverse the bulk $\gmod = 0$.

Overlap Hamiltonians with larger $k$ (corresponding to better quantization) need larger volumes to accurately reproduce the anomaly.
In \cref{fig:anomaly-inflow}(b), we show the vacuum chiral charge $\braket< \Qa(A) >$ for various $k$.
The linear behavior predicted by \cref{eq:Q(A)} is well satisfied for small $k$, even at modest volumes of $L=24$.
However, larger values of $k$ deviate significantly from the linear relationship unless $L$ is increased correspondingly, as can be seen for $L = 128$.
Indeed, \cref{fig:anomaly-inflow}(a) reveals that as $k$ increases, only modes for which $\gmod$ changes rapidly can contribute to the linear dependence on $A$ in \cref{eq:Q(A)}.
At fixed volume $L$, increasing $k$ leads to better charge quantization but reduces the number of such modes, exacerbating finite-size effects.
Conversely, increasing $L$ with fixed $k$ increases the number of modes in regions with rapidly varying $\gmod$, improving convergence to the linear relationship prescribed by \cref{eq:Q(A)}.
The infinite-volume $L \to \infty$ and quantized chiral charge $k \to \infty$  limits fail to commute.
Instead, the anomaly necessitates a simultaneous limit $k, L \to \infty$ with a fixed ratio $k/L$.

Indeed, any lattice formulation with exactly quantized chiral charge that commutes with an onsite vector charge can reproduce the anomaly only in the infinite-volume limit.
In fact, \cref{eq:Q(A)} already implies that a quantized definition of $\Qa(A)$ that commutes with the Hamiltonian cannot exist, since the right-hand side varies continuously with $A$.
This constitutes another simple proof of a
no-go theorem similar to Ref.~\cite{fidkowski_nogo_2023}, showing that a quantized chiral charge that commutes with an onsite $U(1)$ vector symmetry cannot exist.

To quantify the finite-size effects, we examine the vacuum expectation value
\begin{align}
  \Canom(A,L) = \frac{\pi}{L} \ab< \pdv{\Qa}{A}>
  \label{eq:anomaly-C}
\end{align}
for the $k$-overlap Hamiltonians with the background gauge field $A$.
The continuum prediction from \cref{eq:Q(A)} yields $\Canom(A,L) \to 1$ as $L\to \infty$ for any $A$.
\Cref{fig:anomaly-inflow}(c) demonstrates how $k$-overlap Hamiltonians satisfy \cref{eq:anomaly-C}.
The convergence is exponential, $\Canom(A,L) = 1 + \Order(e^{- L / \xi_{k}})$, with the decay length $\xi_{k}$ increasing linearly with $k$,
as illustrated in the inset of \cref{fig:anomaly-inflow}(c) where $\xi_{k} \approx 1.6k$.
This is in accordance with the estimate in \cref{eq:Hchi-decay} and approaches a power law decay as $k \to \infty$.

\section{Conclusions}
\label{sec:conclusions}

A Hamiltonian perspective can yield new insights into the problem of constructing a practical lattice formulation of chiral gauge theories.
Even though the problem of chiral gauge theories remains unsolved, a Hamiltonian \ac{GW} framework is desirable, as it enables the use of quantum computing and tensor-network methods for theories like \ac{QCD} that exhibit a global chiral symmetry with a 't Hooft anomaly.

Although there are obstructions to analytically continuing the Euclidean overlap operator into Minkowski spacetime \cite{mandula_symmetries_2009},
Refs.~\cite{creutz_new_2002,cheluvaraja_axial_2001} proposed an ansatz for an overlap Hamiltonian with an exactly conserved $U(1)$ chiral symmetry that reproduces the anomaly.
However, the chiral-charge operator thus obtained does not have an integer spectrum and therefore does not generate a compact $U(1)$ symmetry.
Motivated by the connection of the standard Hamiltonian overlap \cite{creutz_new_2002, cheluvaraja_axial_2001} to a higher-dimensional domain-wall operator \cite{creutz_surface_1994}, we investigated how the chiral symmetry of the standard overlap Hamiltonian can be systematically improved.
We proposed replacing the conventional \ac{GW} relation in the standard overlap Hamiltonian with a higher-order generalization of the \ac{GW} relation.
This led us to a new family of order-$k$ overlap Hamiltonians with exactly conserved chiral charges for each $k \geq 0$.
Interestingly, the limit $k \to \infty$ is well-defined and yields an overlap operator with an integer chiral charge.
While increasing $k$ improves the quantization of the chiral charge, it comes at the cost of increased nonlocality.
This trade-off, which is inevitable given the no-go theorem \cite{fidkowski_nogo_2023}, is made manifest by the parameter $k$.

Recently, attention has been devoted to the challenges of formulating an exact chiral symmetry in the Hamiltonian formulation, following the no-go theorem of Ref.~\cite{fidkowski_nogo_2023}.
In particular, Ref.~\cite{clancy_ginspargwilson_2024} proposed a definition of the overlap Hamiltonian with a quantized $\gamma_{5}$, but it requires ghost fields with zero chiral charge. This is analogous to our $k \to \infty$, and thus the ghost fields can be interpreted as our bulk modes.
On the other hand, recently, Ref.
\cite{chatterjee_quantized_2024} (and earlier works \cite{horvath_ginspargwilson_1998,horvath_ginspargwilson_1999})
pointed out that there is, in fact, a quantized charge for the $1+1$-dimensional staggered fermion.
(Ref.~\cite{yamaoka_quantized_2025} used a Wilson formulation for the quantized chiral charge.)
It avoids our no-go argument since the quantized chiral charge does not commute with the Hamiltonian in the presence of background $U(1)$ gauge fields, allowing the vacuum expectation value $\braket< \Qa(A) >$  to be noninteger as required by \cref{eq:Q(A)} even if the operator $\Qa$ has a quantized spectrum.
Ref.~\cite{gioia_exact_2025} also proposed a Hamiltonian with exact non-onsite, unquantized chiral symmetry, with properties similar to the overlap Hamiltonians considered in this work.
It is an interesting question whether these recent constructions can be incorporated within a \ac{GW} framework.
In Ref.~\cite{haegeman_interacting_2024}, the authors advocated the use of tangent fermions \cite{stacey_eliminating_1982, stacey_spectrum_1985} which have a quantized charge but a nonlocal Hamiltonian.
Although, as we have argued, such a formulation can only strictly reproduce the anomaly in an infinite volume, tensor-network methods that operate directly in an infinite volume may provide an interesting approach to circumvent this limitation.

A key feature of higher-order $k$-overlap Hamiltonians is their connection to a higher-dimensional domain-wall setup, which
provides physical insight into the locality-quantization tradeoff.
In the domain-wall picture, the (normalized) displacement of single-particle modes from the bulk center measures the chiral charge.
This observation clarifies the seemingly mysterious $\gmod \sim 0$ modes present in the overlap Hamiltonians: these are precisely the bulk modes responsible for enabling anomaly inflow.
We cannot expect to decouple them unless all 't Hooft anomalies cancel.
Instead, as $k$ increases, the $k$-overlap Hamiltonians improve chirality by separating modes either to the wall ($\gmod=\pm 1$) or to the bulk ($\gmod = 0$).

This family of generalized $k$-overlap Hamiltonians lends itself to quantum computing and tensor network methods.
Despite the increased nonlocality, the enhanced chiral properties could make this trade-off advantageous for applications where chiral symmetry plays a crucial role.
In particular, the $k=0$ and $k=\infty$ overlap Hamiltonians represent opposite extremes in the locality-quantization trade-off.
It would be interesting to conduct a systematic study of the resource requirements for these limiting cases on quantum computing platforms.

\section*{Acknowledgements}

I would like to thank  
Aleksey Cherman,
David Kaplan,
Henry Lamm,
Laurens Lootens,
Mendel Nguyen,
Srimoyee Sen,
Shu-Heng Shao,
and
Frank Verstraete
for numerous interesting conversations.
I extend my gratitude to Henry Lamm for a careful reading of this manuscript and for providing valuable feedback.
This work is supported by the Department of Energy through the Fermilab QuantiSED program in the area of ``Intersections of QIS and Theoretical Particle Physics.''
This manuscript has been authored by the Fermi Forward Discovery Group, LLC, under Contract No. 89243024CSC000002 with the U.S. Department of Energy, Office of Science, Office of High Energy Physics.

\bibliography{refs}

\appendix

\section{Explicit construction of higher-order overlap Hamiltonians}
\label[appendix]{sec:explicit}

Explicit construction of the $k$-overlap Hamiltonians can be given in analogy with Fujikawa's construction of the higher-order Euclidean overlap operators \cite{fujikawa_algebraic_2000}.
Let $n \equiv 2k+1$.
First, we note that the $n$th power of the order-$n$ Hamiltonian overlap $\hov[n] \equiv \hov^{n}$ satisfies the standard ($k=0$) \ac{GW} relation.
This suggests that an overlap-like solution for $\hov[n]$:
\begin{align}
	\hov[n] &= \gamma_0 \frac{1 + \Vov[n]}{2}, \quad
  \gmod = \gamma_{5} \frac{1-\Vov[n]}{2},
\end{align}
for some operator $\Vov[n]$.
Requiring $\hov[n]$ to be Hermitian restricts $\Vov[n]$ to be unitary.
Assuming a suitable $\Vov[n]$ can be found, the order-$n$ overlap Hamiltonian $\hov$ can be defined by taking the $n$th root:
$\hov \equiv \ab[\hov[n]]^{\frac{1}{n}} $.
Since $h_{n}$ is Hermitian, all eigenvalues are real, and the $\frac{1}{n}$ root is well-defined in the basis of eigenstates of $\hov[n]$.
Let $\ket| \lambda >$ be an eigenstate of $\hov[n]$ such that $\hov[n] \ket| \lambda > = \lambda \ket| \lambda >$.
The $k$-overlap Hamiltonian is
\begin{align}
  \hov  = \ab[\hov[n]]^{\frac{1}{n}}  \equiv \sum_{\lambda} \sign(\lambda) |\lambda|^{\frac{1}{n}} \ket| \lambda > \bra< \lambda |.
\end{align}
Now, all we need is an explicit form for $\Vov[n]$.
We first define
\begin{align}
	\Dw[n] &= i (\hat\deltasl)^{n} - \ab( r \Delta / 2 )^{n} - m^{n} \label{eq:Dw-n} \\
  i \hat{\delta}_{a} &= \frac{i}{2} ( \delta_{a} - \delta_a^{\dagger}), \quad
        \dsh = {\tsum_{a=1}^{d} \gamma^a \hat{\delta}_{a}} \\
  \Delta &= \tsum_{a=1}^d (\delta_{a} + \delta_{a}^{\dagger})
\end{align}
where
with $\delta_{a}$ being the forward-difference operator acting as
$ \delta_{a} \psi_{x} =   \psi_{x + \hat{a}} - \psi_{x}$,
for $\hat{a}$ a unit vector in the spatial direction $a = 1, \dotsc, d$.
For $n=1$, this is the standard Wilson-Dirac operator with the parameters $m,r$.
Note that $i\dsh$ is anti-hermitian with our convention of anti-hermitian $\gamma_{a}$ matrices.
In momentum space, this is
\begin{align}
  \Dw[n](p) &=
              \ab( \slashed{\hat{p}} )^{n} + \ab[ r B(p) ]^{n} - m^{n}
\end{align}
with $B(p) = \tsum_{a=1}^{d}1 - \cos( p_{a})$,
where we use $\hat{p}_{a} = \sin(p_{a})$ for brevity and $\slashed{\hat{p}} = \sum_{a=1}^{d}\gamma^a \sin(p_{a})$.
We define the unitary operators $\Vov[n]$, in analogy with the standard overlap,
\begin{align}
	\Vov[n] &= \varepsilon( \Dw[n] ) \\
  \Vov[n](p)
        &= \frac{ \slashed{\hat{p}}^{n} + \ab[r B(p)]^{n} - m^{n} }{ \sqrt{ \ab[ (rB(p))^{n} - m^n ]^{2} + (\hat{p}^{2})^{n}  }}
\end{align}
where $\varepsilon(X) = X / \sqrt{{X}^{\dagger} X}$.
As in the standard overlap, we set $0 < m < 2r$ to avoid doublers, and we further require $2m^{2k+1} = 1$ for correct normalization of the low-energy limit of the $k$-overlap Hamiltonian.
In this work, we therefore set $r = 1$ and $m = 2^{-1/ (2k+1)}$ for all numerical computations.
Note that the $U(1)$ vector symmetry is onsite and therefore can be gauged in a straightforward way.
This completes the construction of the $k$-overlap Hamiltonians.

We also note that closed-form expression for the Hamiltonian $k$-overlap can be written in momentum space
\cite{fujikawa_locality_2001, chiu_can_2001, chiu_ginspargwilson_2001},
\begin{align}
	h(p) = \gamma_0 \ab(\frac{1}{ 2 f_{k} } )^{a_{k}}
  \ab\{ \ab(f_k + m_{k})^{a_{k}} - \ab(f_{k} - m_{k})^{b_{k}} \hat{\slashed{p}}  \}
\end{align}
with
\begin{align}
	f_{k}(p) &= \sqrt{(\hat{p}^{2})^{2k+1} + m_{k}^{2}}\\
  m_k(p) &= B(p)^{2k+1} - m^{2k+1}\\
  a_k &= \frac{{k+1}}{2k+1}, \quad
  b_{k} = \frac{k}{2k+1}.
\end{align}

Finally, for the analysis of finite-size effects considered in this work, we note that the quantity $\Canom(A,L)$ defined in \cref{eq:anomaly-C} can be written as
\begin{align}
	  \Canom(A,L) = \frac{\pi}{L} \ab< \pdv{\Qa}{A}> = -\frac{\pi}{2L} \ab< {\psi}^{\dagger} \gamma^5 \pdv{\Vov[n]}{A} \psi >,
\end{align}
which is how we numerically compute $\Canom(A,L)$ for \cref{fig:anomaly-inflow}.


\end{document}